\documentclass[fleqn,usenatbib]{mnras}

\usepackage[T1]{fontenc}

\DeclareRobustCommand{\VAN}[3]{#2}
\let\VANthebibliography\thebibliography
\def\thebibliography{\DeclareRobustCommand{\VAN}[3]{##3}\VANthebibliography}

\usepackage{graphicx}	% Including figure files
\usepackage{amsmath}	% Advanced maths commands
\usepackage{amssymb}	% Extra maths symbols
\usepackage{newtxtext,newtxmath}

\title[FUV Insights into NGC 1399's Globular Clusters]{Far Ultra-Violet Insights Into NGC 1399's Globular Cluster Population}

\author[K. C. Dage et al.]{
Kristen C. Dage,$^{1,2}$\thanks{E-mail: kristen.dage@mcgill.ca}, Yifan Sun$^{1,2}$,  Arunav Kundu$^{4}$, Stephen E. Zepf$^{3}$, Daryl Haggard$^{1,2}$
\\
$^{1}$Department of Physics, McGill University, 3600 University Street, Montr\'eal, QC H3A 2T8, Canada\\
$^{2}$McGill Space Institute, McGill University, 3550 University Street, Montr\'eal, QC H3A 2A7, Canada\\
$^{3}$Department  of  Physics  and  Astronomy,  Michigan  State  University,  East Lansing, MI 48824, USA\\
$^{4}$Eureka Scientific, Inc., 2452 Delmer Street, Suite 100 Oakland, CA 94602, USA\\
}

\date{Accepted XXX. Received YYY; in original form ZZZ}

\pubyear{2022}

\begin{document}
\label{firstpage}
\pagerange{\pageref{firstpage}--\pageref{lastpage}}
\maketitle

\begin{abstract}
We investigate archival Hubble Space Telescope ACS/SBC F140LP observations of NGC~1399 to search for evidence of multiple stellar populations in extragalactic globular clusters. Enhanced FUV populations are thought to be indicators of He-enhanced second generation populations in globular clusters, specifically extreme/blue horizontal branch stars. Out of 149 globular clusters in the field of view, 58 have far ultraviolet (FUV) counterparts with magnitudes brighter than 28.5. Six of these FUV-deteced globular clusters are also detected in X-rays, including one ultraluminous X-ray source ($L_X > 10^{39}$ erg/s). While optically bright clusters corresponded to brighter FUV counterparts, we observe FUV emission from both metal-rich and metal-poor clusters, which implies that the FUV excess is not dependent on optical colour. We also find no evidence that the cluster size influences the FUV emission. The clusters with X-ray emission are not unusually FUV bright, which suggests that even the ultraluminous X-ray source does not provide significant FUV contributions.  NGC 1399 is only the fourth galaxy to have its globular cluster system probed for evidence of FUV-enhanced populations, and we compare these clusters to previous studies of the Milky Way, M31, M87, and the brightest cluster in M81. These sources indicate that many globular clusters likely host extreme HB stars and/or second generation stars, and highlight the need for more complete FUV observations of extragalactic globular cluster systems. %!!! like the abstract. !!!
\end{abstract}


%%%%%%%%%%%%%%%%%%%%%%%%%%%%%%%%%%%%%%%%%%%%%%%%%%

%%%%%%%%%%%%%%%%% BODY OF PAPER %%%%%%%%%%%%%%%%%%

\section{Introduction}
Globular clusters (GCs) have long been test-beds for understanding stellar populations, with implications for many areas of Galactic and extragalactic astronomy. Far ultra-violet (FUV) photometry of globular clusters provides unique constraints as a powerful probe of helium enhanced second populations of stars. Studies by \cite{Catelan09,Dalessandro12} suggest that the FUV emission is likely produced by hot populations of Helium core burning [extreme] horizontal branch (HB) stars, rather than the cool older populations.As shown by \cite{Lee05}, the surface temperatures become extremely hot because the extreme HB stars have a very thin envelope, and thus a He enhanced population will have bluer temperatures.

While many mechanisms have been suggested to explain the presence of multiple stellar populations in globular clusters (see \citealt{Peacock18} and references therein for a summary of many of the proposed scenarios), there is currently no one theory that can explain all of the observational signatures \citep{Bastian18}. Clusters with second generation stars are thought to have undergone multiple star formation episodes, however, the source of the gas is unknown. Excitingly, the presence of stars embedded in gas also has implications as a potential formation channel for gravitational wave mergers \citep[e.g.][]{Rozner22}. 

He-core burning HB branch stars may also explain why some early-type galaxies are much brighter in FUV than expected \citep{Code1969, Dorman95, OConnell99, Brown2000}. \cite{Sandage1960} proposed that metallicity impacts the nature of the HB stars, as metal rich clusters will produce more red HB stars (the `first parameter'). However other physical properties of the cluster also
appear to affect/modulate the structure of the HB/extreme HB \citep{Bellazzini01}. Various second parameters have been
proposed, such as age, helium abundance, stellar core rotation and globular cluster core density \citep{FusiPecci97, Catelan09}. 

Multiple stellar populations have been found in most Galactic globular clusters \citep[and many references therein]{Gratton12}. The much larger population of extragalactic globular clusters also show evidence for second generation stars, as a significant number of clusters studied in M81 \citep{Mayya13}, M87 \citep{Sohn06, Peacock17}, and M31 \citep{Peacock18} have enhanced FUV emission. Spatially resolved analysis of M31 clusters by \cite{Peacock18} finds that the FUV bright sub-population is spatially homogeneous, i.e., the HB stars are not dynamically enhanced.

\cite{Peacock17} found that M87's globular cluster systems show a FUV excess that cannot be solely attributed to red leak, and that even clusters with higher metallicities produce a FUV excess. This suggests that the clusters host He-enhanced second generation population stars$--$even at high metallicities, He-enhanced populations can produce bluer horizontal branch stars. However, to date, only a small sample of metal-rich clusters have been observed in the FUV, and a larger population of metal-rich clusters is needed to demonstrate the distribution of He abundance and ubiquity of second generation stars. 

\begin{figure*}
\begin{tabular}{ll}
\includegraphics[width=3.2in]{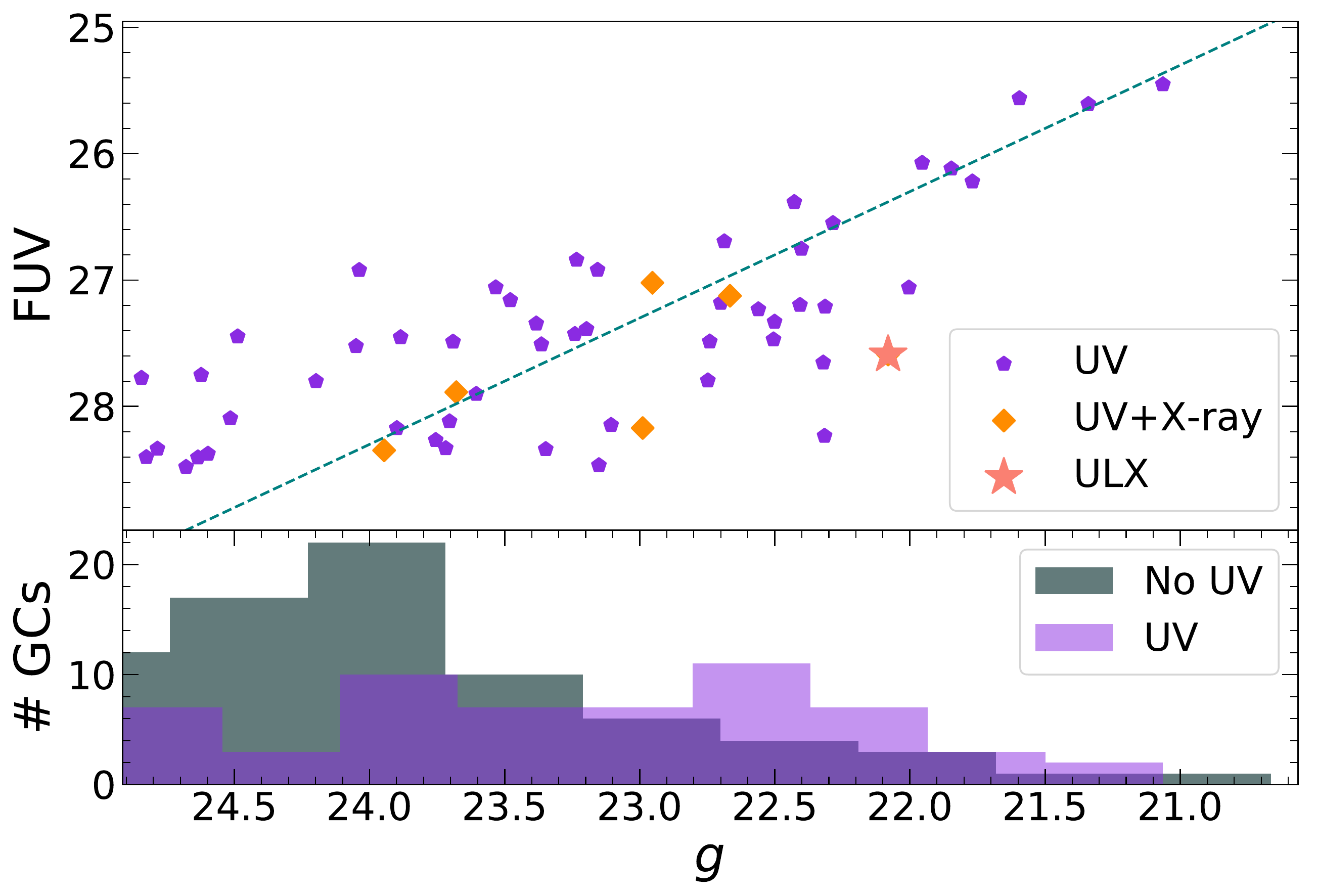}
%&
\includegraphics[width=3.2in]{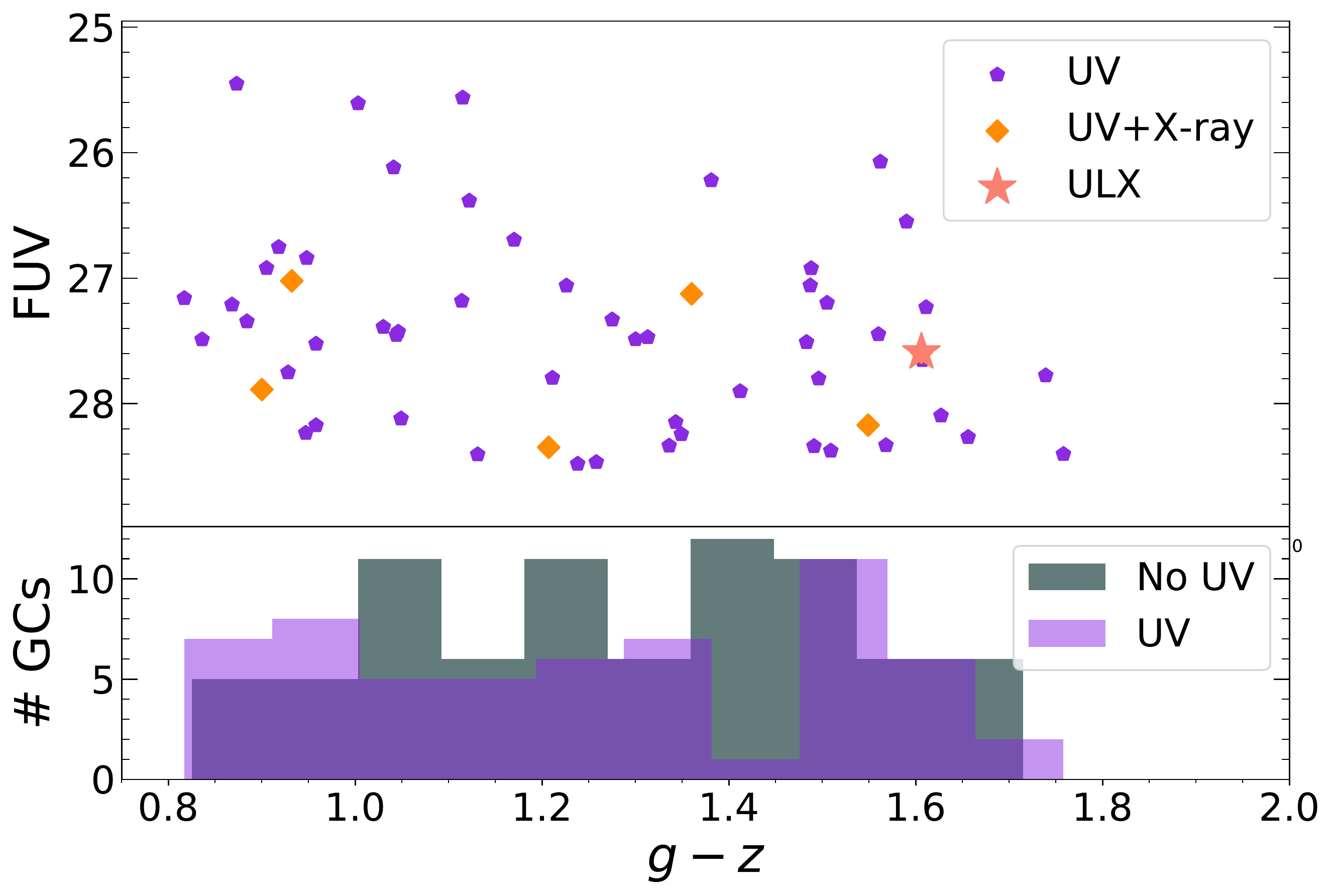}
\end{tabular}
\caption{\textbf{Left:} FUV magnitude versus $g$ for NGC 1399's globular clusters, with a 1:1 correlation line overplotted to guide the eye.  \textbf{Right:} FUV magnitude versus colour ($g-z$) for NGC 1399's globular clusters. There is no apparent correlation between FUV magnitude and colour. The lower panels show a histogram of NGC 1399 clusters versus $g$ or $g-z$, with and without FUV counterparts. We cross-matched the globular clusters with the Chandra Source Catalogue and found six X-ray counterparts, including an ultraluminous X-ray source. } 
\label{Fig:N1399_g_gz}
\end{figure*}

Deep FUV (F140LP) observations of three fields of NGC~1399 were taken with HST ACS/SBC on 2007-08-21 (Field 1), 2007-08-23 (Field 3), and 2008-01-18 (Field 2).These fields contain close to 150 clusters identified by \cite{Jordan15}. We leverage this existing data set to identify more FUV bright globular clusters. Section \ref{sec:data} describes the data-set in detail, along with the analysis process. In Section \ref{sec:results}, we compare the NGC 1399 clusters to the previously studied clusters, along with predictions from stellar synthesis population models. We discuss the implications of this study in Section \ref{sec:discussion}. 

\section{Data and Analysis}

\label{sec:data}
We analysed FUV observations from HST and compared to globular cluster catalogues from \cite{Jordan15}, as well as cross-matching to the Chandra Source Catalogue 2.0 \cite{Evans}. The data and analysis methods are described below.

\subsection{Ultraviolet and optical data}
Three fields of NGC~1399 has been observed with \textit{Hubble Space Telescope}'s Advanced Camera for Surveys (ACS) Solar Blind Channel (SBC) with the F140LP FUV filter, on 2007-08-21 (Field 1), 2007-08-23 (Field 3), and 2008-01-18 (Field 2), proposal ID 10901. Each field was observed for a total of 10800 seconds. We used the drizzled images provided by the Hubble Legacy Archive for this study. We identified any point sources using \textsc{SourceExtractor} \citep{Bertin1996}, as well as to compute their magnitudes,  with a aperture diameter of 5 pixels (0.16''), background of 3 pixels (0.1''), and AB zero-point=23.413\footnote{\url{https://acszeropoints.stsci.edu/}}. 

\begin{figure}
    \centering
    \includegraphics[width=3.5in]{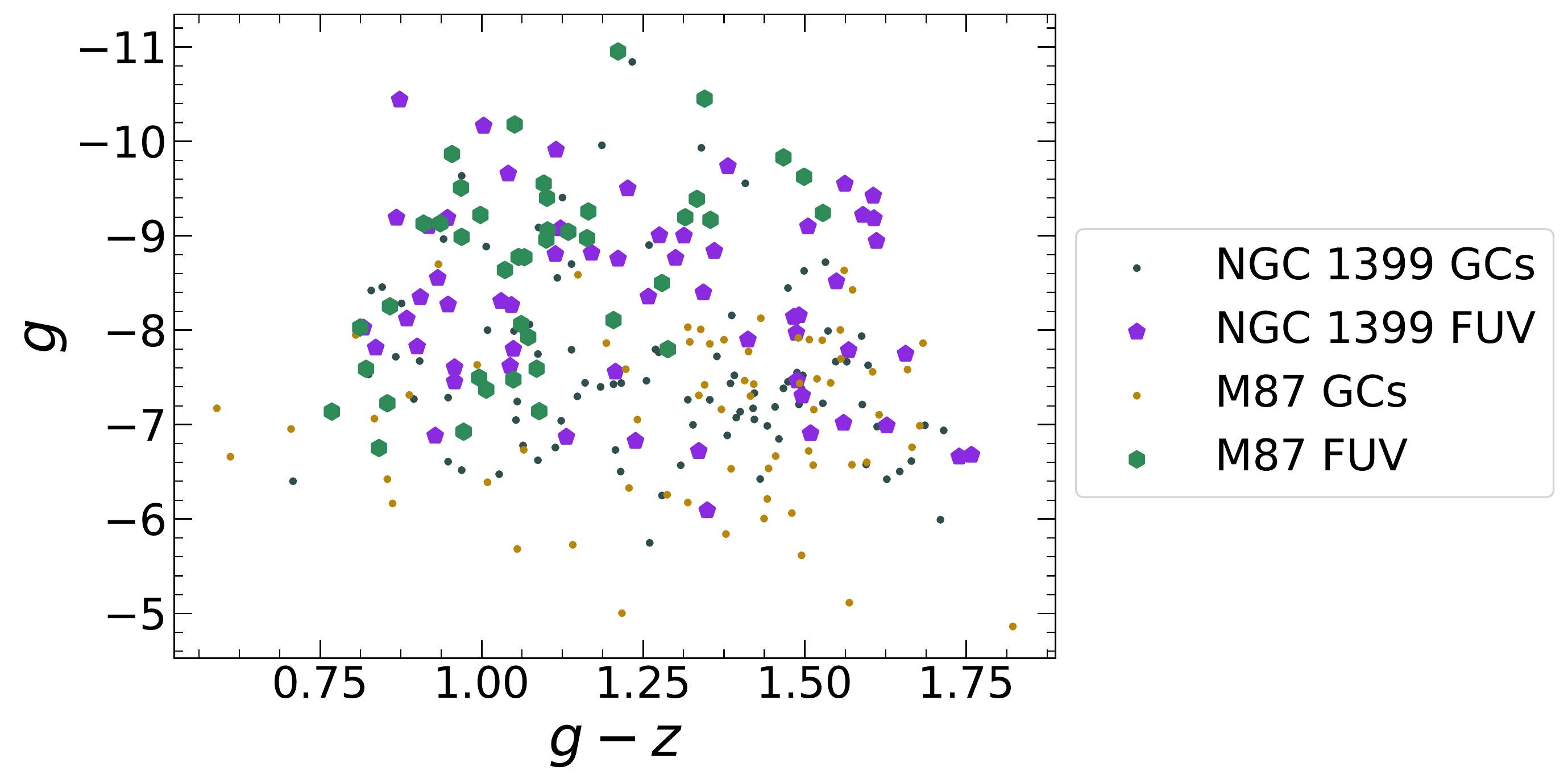}
    \caption{$g$ versus $g-z$ for M87 \citep{Sohn06} and NGC 1399's globular clusters. }
    \label{fig:g_gz_m87_1399}
\end{figure}

We cross-matched the FUV detections to globular cluster candidates identified by the ACS Fornax Cluster Survey \citep{Jordan15} using a 1" tolerance. The three fields cover 149 globular clusters, and of these, 58 are detected in FUV with magnitudes greater than 28.5. Figure \ref{Fig:N1399_g_gz} (right panel) shows the range of FUV and g magnitudes of these clusters, along with the number of clusters with and without FUV counterparts. Clusters with optical magnitudes brigher than g=23.5 appear to have FUV counterparts, which implies that detection of FUV sources is completeness limited. The right panel of Figure \ref{Fig:N1399_g_gz} shows a similar plot, but with colour ($g-z$). This suggests that while brighter clusters are more likely to have FUV counterparts, a wide range of cluster metallicities can be detected in FUV, with no apparent correlation between cluster colour and FUV magnitude. The optical properties of NGC 1399 and M87 clusters with and without FUV emission are displayed in Figure \ref{fig:g_gz_m87_1399}.

\begin{figure}
    \centering
    \includegraphics[width=3in]{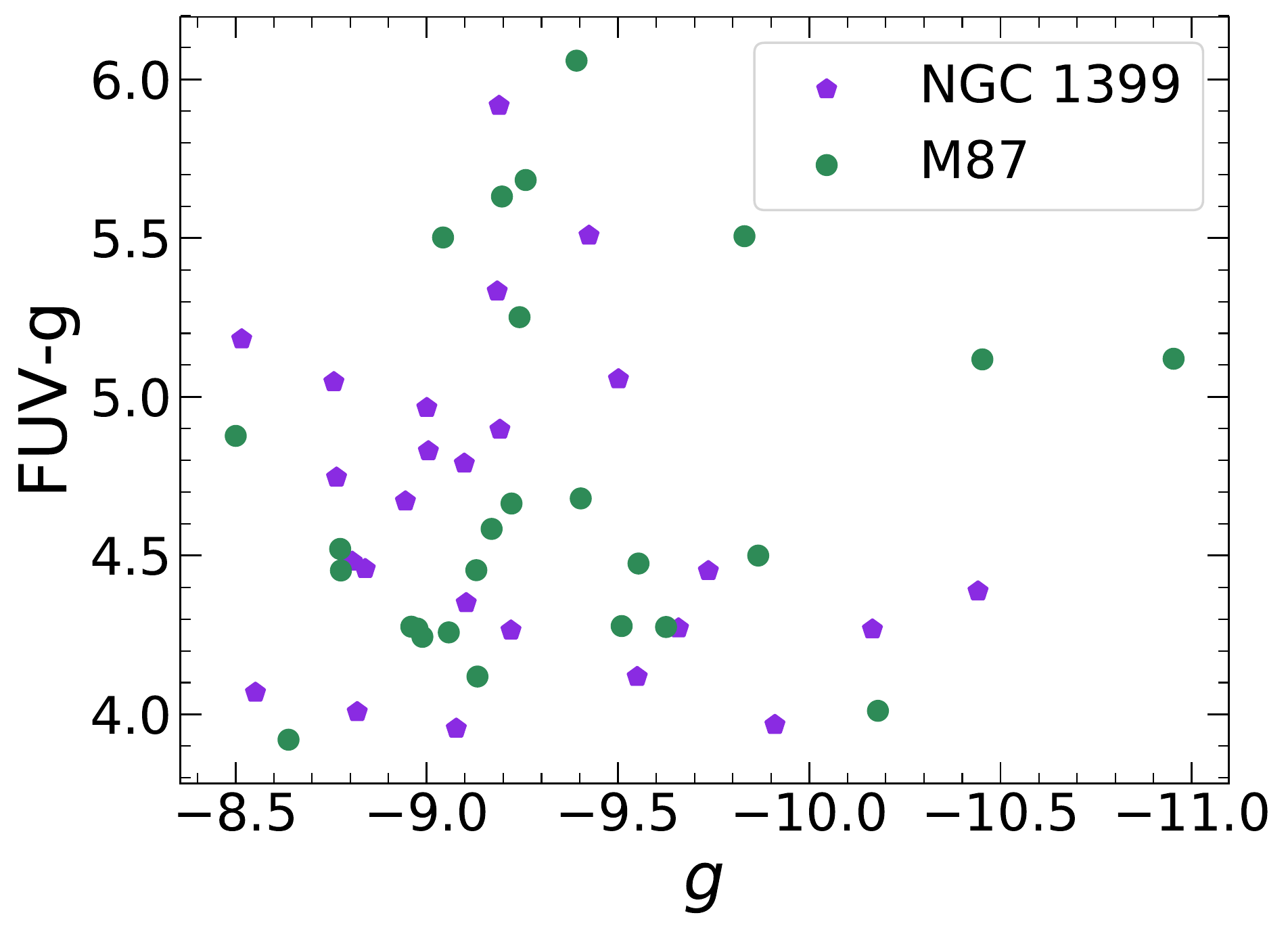}
    \caption{FUV-$g$ versus $g$ for NGC 1399 (this work, \citealt{Jordan15}) and M87 \citep{Sohn06, Jordan09}. }
    \label{fig:fuv-g}
\end{figure}

\begin{figure}
    \centering
    \includegraphics[width=3in]{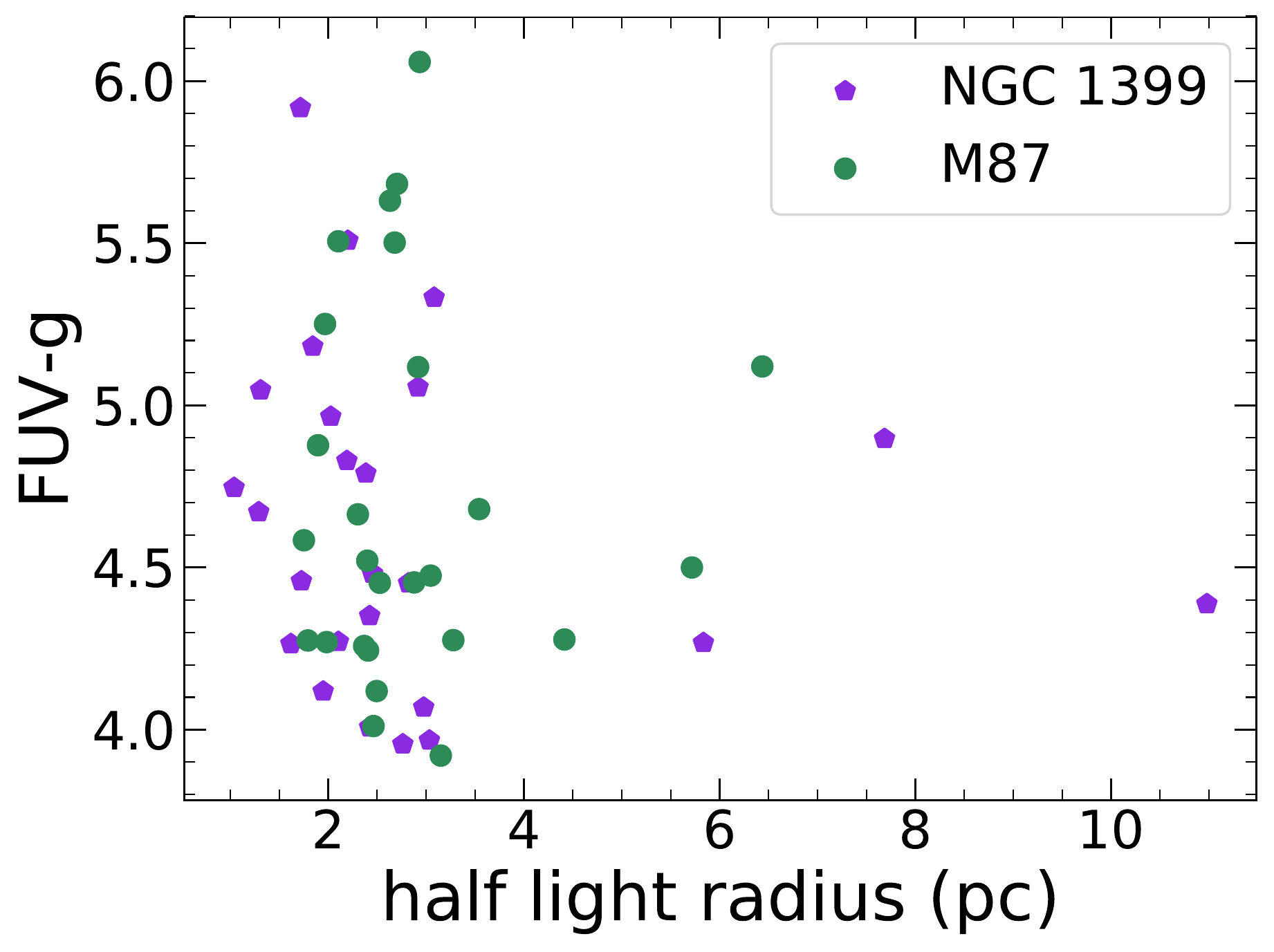}
    \caption{FUV-$g$ versus half-light radius for NGC 1399 (this work, \citealt{Jordan15}) and M87 \citep{Sohn06, Jordan09}. }
    \label{fig:fuv-rh}
\end{figure}

Comparing the FUV-$g$ to $g$ of M87 and NGC 1399's clusters (Figure \ref{fig:fuv-g}), we again do not find any evidence for a redder colour in FUV-$g$ as a function of cluster optical magnitude. Finally, using the half-light radii from \cite{Jordan09, Jordan15}, we do not see any evidence of FUV enhancement as a function of half-light radius (Figure \ref{fig:fuv-rh}, which is consistent with analysis by \cite{Peacock18}. 

%We also crossmatched the sources to the Chandra Source Catalog 2.0 \citep{Evans}. 
%\begin{figure}
 %   \centering
  %  \includegraphics[width=3.5in]{fuv_g_1399.pdf}
   % \caption{Upper panel: FUV magnitude versus $g$ for NGC 1399's globular clusters, with a 1:1 line overplotted. Lower panel: histogram of the number of NGC 1399 clusters and their $g$ magnitudes$--$ both those with and without FUV counterparts. }
    %\label{fig:FUV_g_1399}
%\end{figure}
%\begin{figure}
 %   \centering
 %   \includegraphics[width=3.5in]{fuv_gngcs_1399.pdf}
    %\caption{Upper panel: FUV magnitude versus colour ($g-z$) for NGC 1399's globular clusters. There is no apparent correlation between FUV magnitude and colour. Lower panel: histogram of NGC 1399 clusters colours, with and without FUV counterparts. }
  %  \label{fig:FUV_gz1399}
%\end{figure}

\subsection{X-ray data}
X-ray binaries can produce significant ultraviolet emission \citep[e.g.][]{vanparadijs},  particularly those at high luminosities, such as the microquasar SS433 \citep{Dolan97, waisberg, middleton21}. Given that some globular clusters are known to host ultraluminous X-ray sources \citep[and references therein]{Dage19}: X-ray binaries with luminosities exceeding the Eddington limit, $\sim 10^{39}$ erg/s for a 10\(M_\odot\) black hole, we cross-matched the NGC 1399 globular clusters with the Chandra Source Catalogue 2.0 \citep{Evans} to search for any evidence of FUV enhancement due to X-ray binaries using \textsc{topcat}\footnote{http://www.star.bris.ac.uk/~mbt/topcat/} and a match radius of 1''. We identified six total sources with X-ray counterparts greater than $10^{38}$ erg/s, including one ultraluminous X-ray source (CXOU0338326-35270567; \citealt{Dage19}). As seen in Figure \ref{Fig:N1399_g_gz}, the six sources span a large range in $g$ and $g-z$ space, and are not exclusively limited to the brightest FUV clusters. This suggests that any FUV enhancement by the X-ray binary is minimal. 

\section{Comparison to Previous FUV studies of globular clusters}
\label{sec:results}
\begin{figure*}
    \centering
    \includegraphics[width=5in]{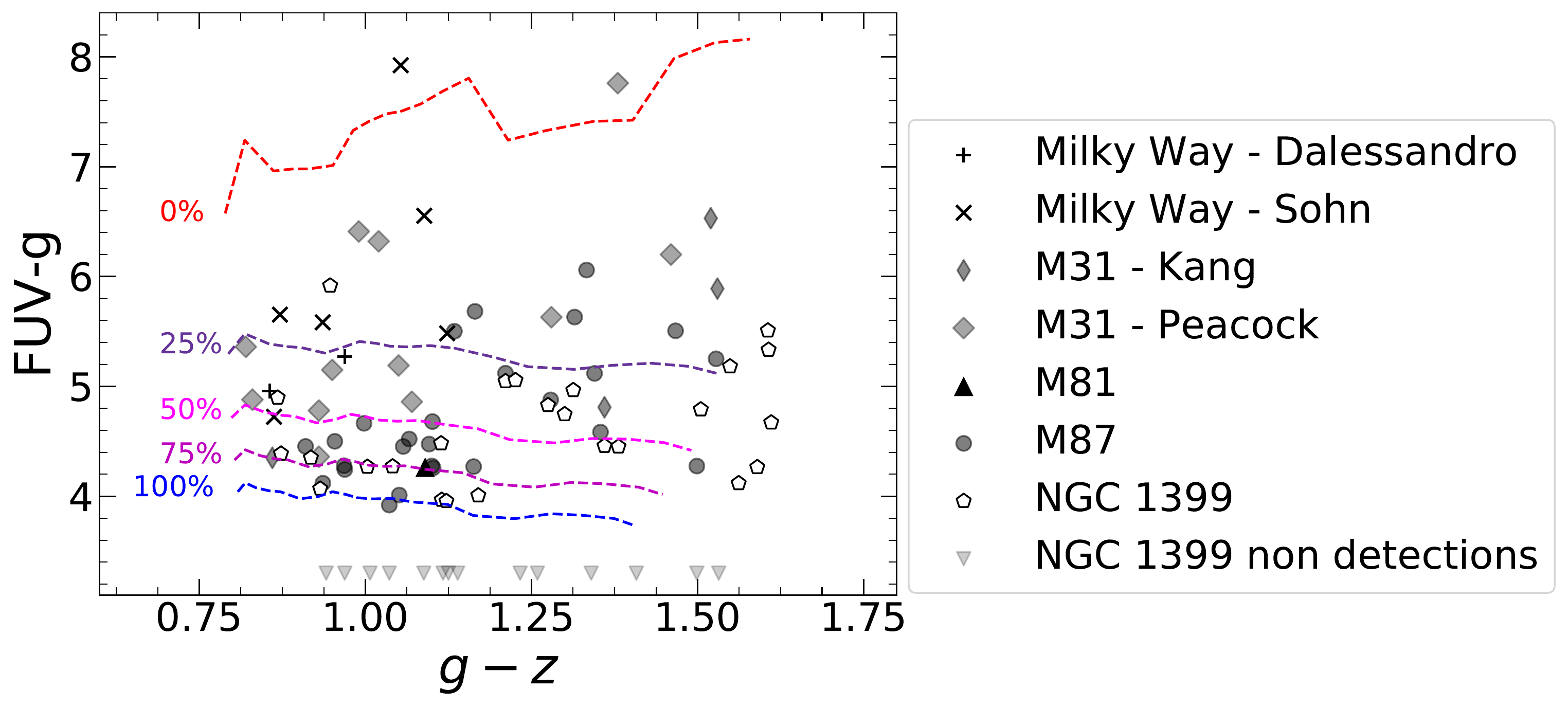}
    \caption{FUV-$g$ versus $g-z$ for the clusters in M31, M81, M87, NGC 1399 and the Milky Way with FSPS models for blue-HB fractions (0\%, 25 \%, 50\%, 75\% and 100\%). The plusses are Milky Way GCs from \citet{Dalessandro12}, crosses are Milky Way clusters from \citet{Sohn06}, small diamonds are M31 clusters from \citet{Kang12}, large diamonds are M31 clusters from \citet{Peacock18}, the triangle is the M81 cluster from \citet{Mayya13}, open circles are M87 clusters from \citet{Sohn06},  and the pentagons are NGC 1399's clusters from this analysis. The triangles show the upper limits of NGC 1399 GCs in the field not detected with FUV. For all clusters, we only select the brightest clusters with $M_g < -8.5$.}
    \label{fig:fsps}
\end{figure*}
FUV enhancement of globular cluster systems have been observed in the Milky Way \citep{Dorman95, Dalessandro12}, M31 \citep{Peacock18}, M81 \citep{Mayya13} and M87 \citep{Sohn06, Peacock17}. Following \cite{Peacock17}, we compare the NGC 1399 globular clusters to the previously studied sample. The archival FUV and optical data for M31, M81, M87 and Milky Way clusters are described below:

\subsection{Milky Way}
% with extinction less than 0.16, as reported by the Harris catalogue \citep{Harris96}. 
We sub-select the brightest ($M_V<-9$) Milky Way globular clusters with FUV (ABmag) measured by \cite{Dalessandro12} and STmag FUV measurements reported by \cite{Sohn06}, originally measured by \cite{Dorman95}. We convert from the STmag to ABmag system by shifting the zero points from 20.629 (as reported in \citealt{Sohn06}) to 23.412.  The  \cite{Dorman95,Sohn06} sample includes two of the most metal rich Galactic globular clusters with FUV measurements: NGC 104 (47 Tuc) and NGC 6441. Due to the nature of Galactic globular clusters, we are restricted to a fairly metal-poor range.

\cite{Dalessandro12} and \cite{Sohn06} report the measurements in FUV-V. We use the relationship from \cite{jester2005} and the reported B-V from the Harris catalogue to convert these measurements from FUV-V to FUV-$g$. We use the relationship from \cite{fahrion2020} to convert the [Fe/H] values from the Harris catalogue to $g-z$.
\subsection{M31}
We use GALEX measurements of M31 clusters reported by \cite{Kang12}. These are already in ABmag. Like before, we sub-select only the brightest ($M_g<-8.5$). We additionally filter on E(B-V), restricting the range to match that of \cite{Peacock18} which we also compare to. \cite{Peacock18} report FUV F140LP measurements of twelve M31 globular clusters in ABmag. All of these clusters are brighter than $M_g$=-8.5 and E(B-V) <0.14. We adopt the $g$ and $z$ magnitudes for these sources from \cite{Peacock10}. 
\subsection{M81}
We use the reported  $g$ and $z$ magnitudes from \cite{Mayya13}'s analysis of the brightest cluster in M81. The FUV measurements are reported in ABmag.

\subsection{M87}
We use the M87 FUV measurements from \cite{Sohn06}, which were reported in STmag. We convert from the STmag to ABmag system like before. Analysis by \cite{Peacock17} suggests that these measurements are 0.5 magnitude too bright due to red leak, and we factor that in. 
We cross match these clusters to \cite{Jordan09}'s cluster catalogue to obtain $g$ and $g-z$ measurements. Given that M87 is the only other distant galaxy with a study of the FUV properties of the extensive GC system, Figure \ref{fig:g_gz_m87_1399} shows a comparison of the optical properties of the two galaxies.

%We use the following relations from \cite{Peacock10} to transform magnitudes from V and V-I to $g$ and $g-z$: V = $g$ - 0.39($g-z$) + 0.07, and $g-z$ = 1.518(V-I) - 0.443. Similar to \cite{Peacock17}, we cut the sample to only include sources with M$_V$ $<$ -9.0. 

\subsection{Synthetic Stellar Population Modeling}
Similar to \cite{Peacock17}, we use the Flexible Stellar Population Synthesis code (FSPS; \citealt{fsps1,fsps2}) to model the hot populations of globular clusters. Following \cite{Peacock17}, we use the PADOVA stellar isochrones and BASEL stellar libraries, with the initial mass function of \cite{chabrier03}, and 
single ages of 13 Gyr. As noted in \cite{Peacock17}, we also shift the TP\textbf{$-$}AGB phase, with \textit{delt} (log $T_{eff}$ of the TP$-$AGB isochrones) = 0.05 and \textit{dell} (log $L_{bol}$ of the TP$-$AGB isochrones)= 0.05 to better replicate M31 and the Milky Way's globular clusters.  The blue-HB star temperatures are uniformly spread in log T$_{eff}$ up to 10,000 K \citep{fsps1}.

The models for fractions of blue-HB stars ranging from 0\% to 100\% are displayed in Figure \ref{fig:fsps}, along with the Milky Way clusters from \cite{Dalessandro12}, M31 \cite{Peacock18}, M81 \cite{Mayya13}, and the brightest clusters from M87 \citep{Sohn06} and NGC 1399. We compare only the brightest clusters corresponding to $M_g$=-8.5, which is where the number of NGC 1399 globular clusters detected in FUV becomes limited. We apply a cut at the same magnitude to the M87 and Milky Way clusters for consistency. The M81 cluster and M31 clusters are all brighter than this limit. While the F140LP observations of NGC 1399 may be susceptible to red leak, work by \cite{Peacock17} suggests that this contribution is at the most $\sim$ 0.5 mag. Such an offset does not significantly impact these results.
\section{Discussion}
\label{sec:discussion}

We analyse HST FUV observations of globular clusters in NGC 1399 and find that a significant number (58 out of 149 clusters) host FUV counterparts, an indicator of extreme/blue HB stars, even in clusters with high metallicity. These stars serve as an indicator of multiple stellar populations in globular clusters, as they are not a typical product of metal-rich environments. Adding to studies of M31, M81, M87 and the Milky Way, this analysis of NGC 1399's globular clusters suggests that He-enhanced populations in globular clusters may be more common than previously thought. Our findings are summarized as follows:

\begin{itemize}
     
    \item Out of 149 globular clusters in the field of view of the FUV observations, 58 are associated with FUV emission brighter than 28.5 magnitude. Both metal-rich and metal-poor clusters are observed to have FUV counterparts, and more optically-bright clusters correspond to more FUV bright counterparts.
    \item The FUV excess does not appear to be dependent on the optical colour. 
    \item In keeping with \cite{Peacock18}, we find no evidence that the spatial size of the cluster influences FUV emission.
      \item While six clusters were detected in both FUV and X-ray, including one ultraluminous X-ray source, there does not appear to be evidence for a significant contribution to excess FUV from X-ray binaries.
    \item This analysis, and other analysis of globular clusters in the Milky Way, M31, M81 and M87 suggests that the contribution of FUV emission from extreme HB stars is far more common than previously thought. 
  
\end{itemize}

NGC 1399 is only the fourth galaxy with an examination of the FUV nature of its globular cluster system. Future deep ultraviolet studies of extragalactic globular cluster systems in ultraviolet, either through continuation of the HST UV-initative, as well as new UV observatories (e.g. ULTRASAT \footnote{\url{https://www.weizmann.ac.il/ultrasat/}} or CASTOR \footnote{\url{https://www.castormission.org/about}}) will be able to probe the ubiquity of enhanced FUV populations and provide stricter constraints on the nature of He-enhanced second generation populations in globular clusters.

\section*{Acknowledgements}
The authors thank the anonymous referee for detailed comments which greatly improved the manuscript. KCD and DH acknowledge funding from the Natural Sciences and Engineering Research Council of Canada (NSERC), and the Canada Research Chairs (CRC) program. KCD acknowledges fellowship funding from the McGill Space Institute and from Fonds de Recherche du Qu\'ebec $-$ Nature et Technologies, Bourses de recherche postdoctorale B3X no. 319864. AK acknowledges funding from grant HST-GO-14738.001-A. SEZ acknowledges funding from grant HST-AR-16160.001-A. This work was performed in part at Aspen Center for Physics, which is supported by National Science Foundation grant PHY-1607611.
%%%%%%%%%%%%%%%%%%%%%%%%%%%%%%%%%%%%%%%%%%%%%%%%%%
\section*{Data Availability}
The HST data is publicly available through the Mikulski Archive for Space Telescopes\footnote{\url{https://mast.stsci.edu/portal/Mashup/Clients/Mast/Portal.html}}.
\newpage
\appendix

\begin{table*}
\caption{Optical and FUV measurements for NGC 1399 globular clusters.Sources marked with $\dagger$ have associated X-ray emission. The optical measurements are from \citet{Jordan15}. }
\begin{tabular}{rrrrrrrr}
\multicolumn{1}{l}{ID}   & \multicolumn{1}{l}{RA} & \multicolumn{1}{l}{Dec} & \multicolumn{1}{l}{F140LP} & \multicolumn{1}{l}{Err} & \multicolumn{1}{l}{$z$} & \multicolumn{1}{l}{$g-z$} & \multicolumn{1}{l}{r$_h$ (pc)} \\ \hline \hline
1                        & 54.6303422             & -35.4575215             & 27.45                    & 0.07           & 22.79                   & 1.04                  & 4.41                  \\
20                       & 54.6380231             & -35.4478644             & 26.12                      & 0.04           & 20.84                   & 1.03                   & 2.10                  \\
26                       & 54.6228213             & -35.4517723             & 26.92                     & 0.06           & 22.59                   & 1.44                    & 3.22                  \\
26                       & 54.6287841             & -35.4415787             & 27.49                      & 0.07         & 22.89                   & 0.82                   & 1.72             \\
31                       & 54.615658              & -35.4502342             & 26.38                      & 0.04          & 21.35                   & 1.09                   & 2.76             \\
31                       & 54.628742              & -35.4424131             & 27.90                       & 0.09          & 22.23                   & 1.36                    & 2.49                   \\
32                       & 54.6190763             & -35.4519047             & 27.51                      & 0.07          & 21.85                   & 1.37                   & 3.42                \\
34                       & 54.6272018             & -35.4428448             & 27.18                       & 0.06          & 21.59                   & 1.08                   & 2.45                  \\
38                       & 54.6245788             & -35.4432121             & 27.49                       & 0.07       & 21.46                   & 1.29                   & 1.04                  \\
39                       & 54.6365662             & -35.44767               & 28.40                       & 0.11    & 23.51                    & 1.15                   & 2.90                     \\
45$\dagger$ & 54.6214564             & -35.4437224             & 27.33                     & 0.07          & 21.27                   & 1.25                   & 2.19                   \\
46                       & 54.6151347             & -35.4524822             & 27.21                      & 0.06           & 21.25                   & 0.87                    & 7.69                \\
48$\dagger$  & 54.6276642             & -35.4440743             & 27.12                       & 0.06          & 21.35                   & 1.33                   & 1.73                \\
49                       & 54.6148802             & -35.4526059             & 27.06                       & 0.06           & 22.05                   & 1.49                   & 2.89                \\
61                       & 54.620959              & -35.4446919             & 27.78                      & 0.08           & 23.10                   & 1.77                 & 2.83                  \\
63                       & 54.61678               & -35.4494839             & 27.23                      & 0.06      & 20.94                   & 1.64                    & 1.29                 \\
76                       & 54.6244563             & -35.4453432             & 25.45                       & 0.03            & 19.85                   & 0.96                  & 10.98                  \\
94                       & 54.6249547             & -35.4462261             & 26.22                  & 0.04          & 20.39                  & 1.36                   & 2.82                   \\
96                       & 54.6224823             & -35.4462521             & 27.80                     & 0.08           & 21.57                   & 1.21                   & 1.31                  \\
99$\dagger$  & 54.6364897             & -35.4481219             & 27.90                       & 0.09          & 22.79                   & 0.89                   & 3.26                 \\
103                      & 54.6256559             & -35.4465556             & 27.20                      & 0.06        & 20.91                  & 1.48                   & 2.38                \\
116 $\dagger$ & 54.6369793             & -35.4496627             & 27.02                      & 0.06           & 22.04                   & 0.90                   & 2.98                  \\
120                      & 54.6232631             & -35.4469452             & 28.17                       & 0.10          & 22.99                   & 0.88                   & 2.22                  \\
136                      & 54.6299692             & -35.4474239             & 27.75                     & 0.08          & 23.59                   & 0.77                  & 3.94                   \\
139                      & 54.6331628             & -35.4499855             & 26.92                      & 0.06          & 22.24                   & 0.91                   & 3.07                  \\
142                      & 54.6334266             & -35.4501091             & 26.55                       & 0.05       & 20.73                    & 1.56                   & 1.62                  \\
142                      & 54.6258497             & -35.4488949             & 28.34                       & 0.11           & 21.88                   & 1.48                   & 1.89                    \\
148                      & 54.6323288             & -35.4503395             & 28.19                       & 0.10           & 22.60                   & 1.10                   & 4.25                   \\
155                      & 54.6296547             & -35.4476052             & 27.06                         & 0.06           & 20.78                   & 1.22                  & 2.92                  \\
158                      & 54.6378408             & -35.4507344             & 26.84                       & 0.05           & 22.28                   & 0.95                  & 3.54                    \\
161                      & 54.6371597             & -35.450862              & 28.15                     & 0.10        & 21.79                   & 1.33                   & 2.19                 \\
166                      & 54.6252193             & -35.4493602             & 27.45                       & 0.07          & 22.83                   & 1.62                   & 5.08                 \\
166                      & 54.6329328             & -35.4504742             & 28.38                        & 0.11          & 23.12                   & 1.49                   & 2.00               \\
176                      & 54.6341175             & -35.4512789             & 28.33                     & 0.11           & 22.17                   & 1.54                   & 1.58                 \\
182                      & 54.6318066             & -35.4514528             & 28.48                       & 0.12         & 23.36                   & 1.29                  & 4.27              \\
184                      & 54.6373224             & -35.4516267             & 25.56                      & 0.03           & 20.49                   & 1.10                   & 3.03                  \\
178$\dagger$ & 54.636168              & -35.4515699             & 27.59                      & 0.08          & 20.49                  & 1.60                     & 2.10                   \\
190                      & 54.6337672             & -35.4516092             & 28.35                    & 0.11            & 22.77                   & 1.18                   & 2.24                 \\
192                      & 54.6288936             & -35.4480532             & 26.70                      & 0.05       & 21.54                  & 1.16                   & 2.42                 \\
216                      & 54.6357635             & -35.4524267             & 27.80                       & 0.08        & 22.69                   & 1.52                   & 3.22                \\
222                      & 54.6316818             & -35.4525526             & 27.39                       & 0.07           & 22.07                   & 1.07                   & 4.78                   \\
227                      & 54.6308562             & -35.4530596             & 25.61                       & 0.03           & 20.28                   & 1.00                   & 5.83                  \\
229                      & 54.634875              & -35.4530886             & 28.17                       & 0.10           & 21.48                   & 1.53                  & 1.84                   \\
242                      & 54.6291993             & -35.453459              & 28.23                      & 0.10            & 21.38                   & 0.97                   & 1.72                 \\
247                      & 54.631239              & -35.4536053             & 27.43                      & 0.07       & 22.20                   & 1.04                  & 4.97                   \\
255                      & 54.6352913             & -35.4538496             & 27.65                       & 0.08        & 20.71                   & 1.61                   & 3.08                   \\
257                      & 54.6322367             & -35.4538418             & 28.47                      & 0.12          & 21.84                   & 1.27                   & 4.27               \\
269                      & 54.6319036             & -35.4543048             & 27.47                      & 0.07          & 21.23                   & 1.29                   & 2.03                  \\
278                      & 54.6315135             & -35.4547057             & 26.75                    & 0.05          & 21.51                   & 0.90                   & 2.42                  \\
281                      & 54.6334479             & -35.4547468             & 26.07                     & 0.04           & 20.43                   & 1.54                   & 1.95                   \\
283                      & 54.6343187             & -35.4547604             & 27.52                      & 0.07           & 23.13                   & 0.94                   & 2.20                  \\
284                      & 54.6375864             & -35.4547734             & 28.24                      & 0.11            & 23.99                   & 1.43                   & 4.08                 \\
288                      & 54.6335195             & -35.4550195             & 27.34                     & 0.07          & 22.50                   & 0.87                   & 2.67                  \\
306                      & 54.6385106             & -35.4559154             & 27.16                      & 0.06        & 22.69                  & 0.79                   & 3.27                   \\
310                      & 54.629859              & -35.4559774             & 28.33                      & 0.11           & 23.44                    & 1.27                & 2.51                   \\
315                      & 54.6337408             & -35.456347              & 28.09                      & 0.10         & 22.93                  & 1.60                    & 2.18                 \\
316                      & 54.6326008             & -35.456412              & 28.27                     & 0.11           & 22.14                    & 1.64                  & 2.82                \\
328                      & 54.6320112             & -35.4569774             & 28.40                     & 0.11         & 23.08               & 1.76                  & 2.33             \\     \hline
\end{tabular}
\end{table*}
 

%\newpage
\bibliographystyle{mnras}
\bibliography{fornaxfuv.bib} % if your bibtex file is called example.bib

%\begin{thebibliography}{99}
%\end{thebibliography}

%%%%%%%%%%%%%%%%%%%%%%%%%%%%%%%%%%%%%%%%%%%%%%%%%%

%%%%%%%%%%%%%%%%% APPENDICES %%%%%%%%%%%%%%%%%%%%%

% Don't change these lines
\bsp	% typesetting comment
\label{lastpage}
\end{document}